\documentstyle[twoside,12pt]{report}

\catcode`\@=11
\long\def\@makefntext#1{ 
\protect\noindent \hbox to 3.2pt {\hskip-.9pt
$^{{\eightrm\@thefnmark}}$\hfil}#1\hfill} 

 \def\@makefnmark{\hbox to 0pt{$^{\@thefnmark}$\hss}}  

\def\ps@myheadings{\let\@mkboth\@gobbletwo
\def\@oddhead{\hbox{} 
\rightmark\hfil\eightrm\thepage}
\def\@oddfoot{}\def\@evenhead{\eightrm\thepage\hfil 
\leftmark\hbox{}}\def\@evenfoot{}
\def\sectionmark##1{}\def\subsectionmark##1{}}

\oddsidemargin=\evensidemargin
\newcounter{sectionc}\newcounter{subsectionc}\newcounter{subsubsectionc}
\renewcommand{\section}[1] {\vspace{12pt}\addtocounter{sectionc}{1}
\setcounter{subsectionc}{0}\setcounter{subsubsectionc}{0}\noindent
	{\bf\thesectionc. #1}\par\vspace{5pt}}
\renewcommand{\subsection}[1] {\vspace{12pt}\addtocounter{subsectionc}{1}
	\setcounter{subsubsectionc}{0}\noindent
	{\bf\thesectionc.\thesubsectionc. {\kern1pt \bf\it #1}}\par\vspace{5pt}}
\renewcommand{\subsubsection}[1] {\vspace{12pt}\addtocounter{subsubsectionc}{1}
	\noindent{\thesectionc.\thesubsectionc.\thesubsubsectionc.
	{\kern1pt \it #1}}\par\vspace{5pt}}

\newcommand{\textlineskip}{\baselineskip=14pt}
\newcommand{\smalllineskip}{\baselineskip=12pt}

\def\eightcirc{
\begin{picture}(0,0)
\put(4.4,1.8){\circle{6.5}}
\end{picture}}
\def\eightcopyright{\eightcirc\kern2.7pt\hbox{\eightrm c}}

\newcounter{itemlistc}
\newcounter{romanlistc}
\newcounter{alphlistc}
\newcounter{arabiclistc}

\newcommand{\fcaption}[1]{
        \addtocounter{figure}{1}
         {{\tenrm Fig.~\thefigure . #1} }\hfil\break }

\newcommand{\tcaption}[1]{			
        \addtocounter{table}{1}
         {{\tenrm\offinterlineskip Table~\thetable . #1} }\hfil\break }

\def\thebibliography#1{\section{References}\list
  {[\arabic{enumi}]}{\settowidth\labelwidth{[#1]}\leftmargin\labelwidth
    \advance\leftmargin\labelsep
    \usecounter{enumi}}
    \def\newblock{\hskip .11em plus .33em minus .07em}
    \sloppy\clubpenalty4000\widowpenalty4000}

\def\pmb#1{\setbox0=\hbox{#1}
	\kern-.025em\copy0\kern-\wd0
	\kern.05em\copy0\kern-\wd0
	\kern-.025em\raise.0433em\box0}

\def\fnt#1#2{\footnotetext{\kern-.3em
	{$^{\mbox{\scriptsize #1}}$}{#2}}}

\def\fpage#1{\begingroup
\voffset=.3in
\thispagestyle{empty}\begin{table}[b]\centerline{\footnotesize #1}
	\end{table}\endgroup}

 1
 1
 1

\font\eightrm=cmr8

\def\qed{\hbox{${\vcenter{\vbox{                          
   \hrule height 0.4pt\hbox{\vrule width 0.4pt height 6pt
   \kern5pt\vrule width 0.4pt}\hrule height 0.4pt}}}$}}

\newcommand{\be}{\begin{eqnarray}}
\newcommand{\ee}{\end{eqnarray}}
\newcommand{\dslash}{\partial \hskip -0.5em /}
\newcommand{\Dslash}{D \hskip -0.7em /}
\newcommand{\Vslash}{V \hskip -0.7em /}

\newcommand{\Aslash}{A \hskip -0.7em /}
\newcommand{\tr}{{\rm tr}}
\newcommand{\Tr}{{\rm Tr}}

\newcommand{\La}{{\cal L}}
\newcommand{\A}{{\cal A}}

\newcommand{\ie}{{\it i.e.}\ }
\newcommand{\eg}{{\it e.g.}\ }
\newcommand{\GK}[1]{\frac{#1}{2G+1}}
\newcommand{\GS}{2 \sqrt{G(G+1)}}
\newcommand{\INT}{N_{nm} \int_{0}^{1}{dx x^2 f(Dx)\ }}

\begin{document}
\normalsize\textlineskip
{\thispagestyle{empty}
\setcounter{page}{1}

\fpage{1}
\rightline{UNIT\"U-THEP-1/1993}
\rightline{March 1993}
\vspace{1cm}
\centerline{\Large \bf The Chiral Soliton of the Nambu--Jona-Lasinio}
\vspace{0.5cm}
\centerline{\Large \bf Model with Vector and Axial Vector Mesons
$^\dagger $}
\vspace{0.37truein}
\centerline{U. Z\"uckert, R. Alkofer, H. Reinhardt
and  H. Weigel}
\vspace{0.37truein}
\centerline{Institute for Theoretical Physics, T\"ubingen University}
\vspace{0.5cm}
\centerline{D-7400 T\"ubingen, FR Germany}

\vspace{4cm}
\normalsize\textlineskip
\noindent
\centerline{\bf Abstract}
\vspace{0.5cm}

\noindent
The self-consistent chiral soliton of the
Nambu--Jona-Lasinio model including the $\omega,\rho$ and
$a_1$ (axial-) vector meson fields besides the chiral angle is
investigated. The resulting energy spectrum of the one particle
Dirac Hamiltonian is strongly distorted leading to a polarized
Dirac sea which carries the complete baryon number. This
supports Witten's conjecture that baryons can be described as
topological solitons. The exploration of the isoscalar mean squared
radius of the nucleon exhibits that the repulsive character of the
isoscalar vector field $\omega$ as well as the attractive features
of the (axial-) vector mesons $\rho$ and $a_1$ are reproduced in the
Nambu--Jona-Lasinio model. The axial charge of the nucleon $g_A$
comes out far too small. This can be understood as an artifact of
the proper time regularization prescription.

\vfill

\noindent
$^\dagger $
{\footnotesize{Supported by the Deutsche Forschungsgemeinschaft (DFG) under
contract Re 856/2-1.}}
\eject

\normalsize\textlineskip
\stepcounter{chapter}
\leftline{\large \bf 1. Introduction}

\medskip

In the last decade a description of baryons as chiral solitons
proved to be quite successful. The soliton picture of baryons is
based on large $N_C$ QCD considerations,  $N_c$ being the number of
colors. In the limit $N_c \to \infty$, QCD is equivalent to an effective
theory of weakly interacting mesons \cite{tHo74}. Later Witten
conjectured that in this effective meson theory baryons emerge as
soliton solutions \cite{Wi79}. Furthermore, in the low-energy limit
this meson theory is dominated by the pseudoscalar mesons, the
would-be Goldstone bosons of spontaneously broken chiral symmetry
described in form of a non-linear $\sigma$-model. In order to
implement the chiral anomaly this non-linear $\sigma$-model has to be
supplemented by the Wess-Zumino action. Introducing external gauge fields
allows to extract the corresponding Noether currents. Especially,
the baryon current arising from the Wess-Zumino term proves to be
identical to the topological current thus supporting Skyrme's
original work \cite{Sky61}.

Incorporating vector and axial vector mesons in the non-linear
$\sigma$-model the Skyrme model arises in the limit of infinitely
heavy vector and axialvector mesons. In addition, the inclusion of
vector and axial vector mesons with their physical masses cures
several deficiencies of the Skyrme model,
as \eg incomplete description of electromagnetic properties due to
 missing vector dominance \cite{Me87}
or wrong ``high energy behavior" of $\pi - N$ phase shifts due to the
higher order stabilization terms\cite{Eck86}. This demonstrates the
important role of vector and axialvector mesons in the soliton description
of baryons.

Although we have good confidence that the non-linear $\sigma$-model
extended by vector mesons represents the low-energy approximation
to the effective meson theory of QCD the actual effective
theory is not explicitly known. Therefore Witten's conjecture cannot
be proven within QCD without using further approximations. On the
other hand, the low-energy form of the effective meson theory is
almost entirely determined by chiral symmetry. Therefore the
soliton picture of baryons should not depend on the details of
chiral QCD dynamics. This suggests to investigate whether Witten's
conjecture is fulfilled in simpler models for the quark flavor
dynamics of QCD. For this purpose the Nambu--Jona-Lasinio (NJL)
model is well suited. First of all, like QCD, it obeys chiral
symmetry and, furthermore, it can be motivated as low-energy
approximation to QCD\cite{Sch90}. Additionally, it can be bosonized
by functional integral methods. The resulting effective meson theory is in
satisfactory agreement with the low-energy meson data. In
fact, its gradient expansion yields in leading order the linear
$\sigma$-model and the Wess-Zumino action.

The bosonized NJL model shows that the topological current arises in
leading order gradient expansion from the vacuum part of the baryon
current. The vacuum is defined with all negative fermion states occupied
and the positive energy states being empty. Therefore the topological
current can describe a non-zero baryon charge only if the vacuum is charged
requiring that the valence quarks are bound into the Dirac sea. Only
then the chiral field can carry a non-trivial baryon number. Thus the key
assumption underlying the soliton picture of baryons is that the
valence quarks have joined the Dirac sea.
Within the NJL model we can test whether this assumption is fulfilled.

In recent years the soliton of the NJL model has been extensively studied.
First calculations were restricted to the chiral field\cite{Re88a,Me89,Al90}.
Extending the model to also include the $\rho$ meson provides only
minor changes to the chiral soliton\cite{Al91}. In both calculations the
energy eigenvalue corresponding to the valence quarks is positive.
The physical picture changes drastically if the chiral partner of the
$\rho$ meson, the $a_1$ axialvector meson, is added. Then the valence
quarks become strongly bound and join the Dirac sea, \ie the Skyrmion
picture of baryons results\cite{Al92}. However, in these calculations
the isoscalar vector meson $\omega$ was left out for technical reasons.
The inclusion of the $\omega$ meson should even favor the Skyrmion picture
since the $\omega$ introduces repulsion and hence increases the spatial
size of the chiral field which in turn increases the binding of the
quarks.

The inclusion of the $\omega$ meson leads to substantial
technical complications because the $\omega$ field
develops a non-zero time-like component in the chiral soliton.
Due to the need for regularization the effective meson theory can be
properly defined only via the continuation to Euclidean space.
Thereby time-components of vector fields are continued to
imaginary values, too. As a consequence the Euclidean action becomes complex
and the evaluation of the NJL action for a soliton field leads to a
non-Hermitean eigenvalue problem, see ref.\cite{Al92a} for a proper
treatment. In this calculation only the $\omega$ meson and the chiral
field have been included in a self-consistent soliton calculation. In
the present paper we present the self-consistent soliton calculation in
the NJL model with all low-lying two flavor vector and axial vector meson
fields included.

\vskip1cm
\stepcounter{chapter}
\leftline{\large \bf 2. Bosonization of the NJL Model}

\medskip

The starting point for the following considerations is
the chirally invariant NJL model \cite{Na61,Eb86}:
\be
\La = \bar q (i\dslash - m^0 ) q  & + & 2g_1 \sum _{i=0}^{N_f^2-1}
\left( (\bar q \frac {\lambda ^i}{2} q )^2
      +(\bar q \frac {\lambda ^i}{2} i\gamma _5 q )^2 \right)
\nonumber \\*
 & - & 2g_2 \sum _{i=0}^{N_f^2-1}
\left( (\bar q \frac {\lambda ^i}{2} \gamma _\mu q )^2
      +(\bar q \frac {\lambda ^i}{2} \gamma _5 \gamma _\mu q )^2 \right) ,
\label{njl}
\ee
wherein $q$ denotes the quark spinors and $m^0$ the current quark mass
matrix.  Here we will work in the isospin limit, \ie $m_u^0=m_d^0=m^0$.
The matrices $\lambda ^i/2$ are the generators of the flavor group
($\lambda ^0 = \sqrt{2/N_f} {\bf 1}$). Furthermore we will restrict
ourselves to two flavors $(N_f=2)$ implying $\lambda ^i =\tau ^i, \,
i=0,\ldots ,3.$ The coupling constants $g_1$ and $g_2$ will be
determined from mesonic properties, cf. \eg refs.\cite{Eb86,Re88,Be88,Kl90}
for the calculation of meson properties in the NJL model.

Applying standard functional integral bosonization techniques the model
(\ref{njl}) can be rewritten in terms of composite meson fields\cite{Eb86}
\be
\A &=& \A _F + \A_m,
\nonumber \\*
\A _F &=& \Tr \log (i\Dslash )
 = \Tr \log \big(i\dslash +\Vslash +\gamma_5\Aslash
- (P_R\Sigma+P_L\Sigma^{\dag}) \big),
\nonumber \\*
\A _m &=& \int d^4x \left( -\frac 1 {4g_1}
\tr (\Sigma^{\dag} \Sigma - m^0(\Sigma+\Sigma^{\dag}
) +(m^0)^2)  + \frac{1}{4g_2}\tr(V_\mu V^\mu+A_\mu A^\mu) \right) .
\nonumber \\*
\label{action}
\ee
Here $P_{R,L} = (1\pm \gamma _5)/2$ are the projectors on right-- and
left--handed quark fields, respectively. $V_\mu=
\sum_{a=0}^3V_\mu^a\tau^a/2$ and $A_\mu=
\sum_{a=0}^3A_\mu^a\tau^a/2$ denote the vector and axial vector meson
fields. $V_\mu^a$ and $A_\mu^a$ are real in Minkowski space. The
complex field $\Sigma$ describes the scalar and pseudoscalar meson fields,
$S_{ij}= S^a\tau^a_{ij}/2$ and $P_{ij}= P^a\tau^a_{ij} /2$:
\be
\Sigma=S + i P = \xi_L^{\dag}\ \Phi\ \xi_R,
\label{MS}
\ee
wherein we already introduced the angular decomposition of the complex
field $\Sigma$ into a Hermitean field
$\Phi$ and unitary fields $\xi_L$ and $\xi_R$ which are related to
the chiral field by $U=\xi_L^{\dag}\xi_R$. The latter is conveniently
expressed in terms of a chiral angle $\Theta $
\be
U(x) = \exp \left( i \Theta (x) \right) .
\label{chfield}
\ee

The quark determinant $\A_F$ diverges and therefore requires regularization.
As in a study of the mesonic sector of the NJL model\cite{Eb86} as well as
in previous studies of the soliton
sector\cite{Re88a,Me89,Al90,Al91} we will use Schwinger's proper time
regularization\cite{Sch51} which introduces an $O(4)$-invariant cut-off
$\Lambda$ after continuation to Euclidean space.
For the regularization procedure it is necessary to consider the real
and imaginary part of $\A_F$ separately
\be
\A_F &=& \A_R+\A_I
\nonumber \\*
\A_R &=& \frac 1 2 \Tr \log ( \Dslash_E ^{\dag}\Dslash_E )
\nonumber \\*
\A_I &=& \frac 1 2 \Tr \log ( (\Dslash_E ^{\dag})^{-1} \Dslash_E ) .
\label{arai}
\ee
The real part $\A_R$ diverges like $\log p^2$
for large momenta $p$ whereas the imaginary part $\A_I$ does not contain
any divergencies, \ie it is finite without regularization. However, we
believe that it has to be regularized also in order to have a consistent model.
After all, the occurrence of the cutoff is a very crude way of mimicing the
asymptotic freedom of QCD.

For the real part the proper time  regularization consists in replacing
the logarithm by a parameter integral
\be
\A_R\longrightarrow - \frac 1 2 \int_{1/\Lambda ^2}^\infty \frac {ds}s
\Tr \exp \left( -s \Dslash_E ^{\dag}\Dslash_E \right) ,
\label{arreg}
\ee
which for $\Lambda \to \infty $ reproduces the logarithm
up to an irrelevant constant. Since the operator $\Dslash_E ^{\dag}
\Dslash_E
$ is Hermitean and positive definite this integral is well defined. For
the imaginary part the regularization procedure is equivalent,
\be
\A_I\longrightarrow - \frac 1 2 \int _{1/\Lambda ^2}^\infty \frac {ds}s \Tr
\exp
\left( -s
(\Dslash_E ^{\dag})^{-1} \Dslash_E \right) ,
\label{aireg}
\ee
however, in this case one has to be careful concerning the convergence
of the integral, see section 3.

In order to determine the coupling constants $g_1$ and $g_2$ from the
meson sector it is sufficient to only inspect $\A_R$.
Varying the regularized effective action with respect to the scalar and
pseudoscalar fields yields the Dyson--Schwinger or gap equations
\be
\langle S _{ij} \rangle &=& \delta _{ij} M
\nonumber \\*
M &=& m^0- 2g_1 \langle \bar q q \rangle
\nonumber \\*
\langle \bar q q \rangle &=& - M^3 \frac {N_c}{4\pi ^2} \Gamma
(-1,M^2/\Lambda ^2) .
\label{gap}
\ee
The quantity $M$ is the dynamically generated
constituent quark mass and $\langle \bar q q \rangle$ the quark
condensate. A non-vanishing quark condensate reflects spontaneous
breaking of chiral symmetry.

Performing the derivative expansion\cite{Eb86} of $\A_R$
allows to read off the pion decay constant $f_\pi$
as the coefficient of the expression quadratic in the derivatives
of the physical pion field $\pi=\sum_{a=1}^3\pi^a\lambda^a$:
\be
\A_R=\int d^4x~ \frac{f^2_\pi}{4}~{\rm tr}\partial_\mu \pi
\partial^\mu \pi +\cdot\cdot\cdot.
\label{derexp}
\ee
The inclusion of vector and axialvector mesons leads to
pseudoscalar--axialvector meson mixing, especially $\pi-a_1-$mixing.
This renormalizes the pion field and thus affects the pion decay
constant which is then given by
\be
f_\pi^2  =  \frac {6M^2}{g_V^2} \frac 1{1+6M^2/m_\rho ^2}
\label{fpi}
\ee
where
\be
g_V = \left( \frac 1 {8\pi ^2} \Gamma (0,\frac {M^2}{\Lambda ^2})
\right) ^{-1/2}
\quad {\rm and} \quad
m_\rho ^2 = \frac {g_V^2}{4g_2}
\label{mrho}
\ee
are the universal vector coupling constant and the vector meson mass.
As input quantities from experiment we use the pion decay constant
$f_\pi=93$MeV and the $\rho$ meson mass $m_\rho=770$MeV. For a given
constituent quark mass $M$ we then determine the cut-off $\Lambda$
via eqn.(\ref{fpi}) and subsequently $g_2$ via (\ref{mrho}).
It is important to note that the $\pi-a_1-$mixing increases
$\Lambda$ significantly, \eg for $M=350$MeV we find
$\Lambda =1274$MeV compared to $\Lambda = 649$MeV when $\pi-a_1-$mixing
is disregarded. Expanding $\A_m$ up to second order in the pseudoscalar
fields allows to express the current quark mass $m^0$ in terms of $g_1$
and the pion mass $m_\pi=135$MeV: $m^0=g_1m_\pi^2f_\pi^2/M$. Finally we
employ the gap equation (\ref{gap}) to eliminate $g_1$ in terms
of the constituent quark mass $M$ which from now on is considered as
the only free parameter of the model.

\vskip1cm
\stepcounter{chapter}
\leftline{\large \bf 3. The Energy Functional for Static Meson Fields}

\medskip

Next we will consider the energy functional of the static soliton in SU(2).
After Wick rotation, \ie $x_0=-ix_4$ and $V_0=-iV_4$ the Euclidean
Dirac operator corresponding to eqn. (\ref{action}) is given by
\be
i\beta \Dslash_E &=& - \partial_\tau-h,
\nonumber \\*
h &=& \mbox {\boldmath $\alpha \cdot p $}+iV_4+i\gamma_5A_4 +
\mbox {\boldmath $\alpha \cdot V $} +
\gamma _5 \mbox {\boldmath $\alpha \cdot A $}
+\beta (P_R\Sigma+P_L\Sigma^{\dag})
\label{euham}
\ee
wherein $\tau$ denotes the Euclidean time. In Euclidean space
$\tau$, $V_4$ and $A_4$ have to be considered as Hermitean quantities.
This leads to a non--Hermitean Hamiltonian $h$ even for static
configurations (\ie $[\partial_\tau,h]=0$) if non--vanishing time
components of vector or axialvector meson fields are included.

We fix the scalar field at its vacuum value $\Phi=M{\bf 1}$.  This
is {\it a priori} an unjustified approximation and indeed it has
been shown\cite{Si92} that allowing $\phi$ to be space dependent leads
to a collapse of the soliton in the case when only scalar and
pseudoscalar fields are present. Employing, however, a well
motivated generalization\cite{Ri92} of the NJL model which includes
an additional $\Phi^4$ potential in the mesonic action (\ref{action})
yields a stable soliton with the scalar field deviating only slightly
from its vacuum value\cite{Wei93,Me92}. Thus it is reasonable to
keep the NJL model(\ref{action},\ref{arai}) and restrict the scalar
field to its vacuum value.

For the chiral field we impose the hedgehog
{\it ansatz}:
\be
U(\mbox {\boldmath $x $})={\rm exp}\Big(i{\mbox{\boldmath $\tau$}}
\cdot{\bf \hat r}\Theta(r)\Big).
\label{ansatz1}
\ee
This configuration has vanishing `grand spin' $\mbox{\boldmath $G$}=
\mbox{\boldmath $l$}+\mbox{\boldmath $\sigma$}/2
+\mbox{\boldmath $\tau$}/2$, \ie $[\mbox{\boldmath $G$},U]=0$.
The only possible ansatz for the isoscalar-vector field $\omega$
with grand spin zero has vanishing spatial components($\omega_i=0$):
\be
V_\mu^0=\omega_\mu=\omega(r)\delta_{\mu4}.
\label{ansatz2}
\ee
Parity invariance requires the isoscalar--axialvector meson field
$A_4$ to vanish in the static limit. For the isovector- (axial) vector
meson fields we use the spherically symmetric {\it ans\"atze}
\be
\quad V_4 ^a&=&0, \quad V_i^a
= \epsilon ^{aki} \hat r^k G(r),
\nonumber \\
\quad A_4^a&=&0, \quad
A_i^a = \hat r^i \hat r^a F(r) + \delta ^{ia} H(r)
\label{ansatz3}
\ee
where the indices $a,i$ and $k$ run from 1 to 3. Note that $V_i^a$
corresponds to the physical $\rho$ meson while the physical axialvector
meson $a_1$ is obtained from $A_i^a$ and terms involving
$\partial_i({\bf \hat r}^a\Theta)$ after removing the pseudoscalar
axialvector mixing. The Euclidean Dirac Hamiltonian now reads
\be
h &=& \mbox {\boldmath $\alpha \cdot p $}+i\omega(r)
+ M \beta({\rm cos}\Theta(r)+i\gamma_5
{\mbox{\boldmath $\tau$}}\cdot{\bf \hat r}{\rm sin}\Theta(r))
\nonumber \\*
&&+\frac 1 2 (\mbox {\boldmath $\alpha $} \times {\bf \hat r} )
{\mbox{\boldmath $\cdot\tau$}} G(r)
+ \frac 1 2 (\mbox {\boldmath $\sigma \cdot $} {\bf \hat r} )
(\mbox {\boldmath $\tau  \cdot $} {\bf \hat r} ) F(r)
+\frac 1 2 (\mbox {\boldmath $\sigma \cdot \tau $} ) H(r)
\label{hamil}
\ee
which obviously is not Hermitean since $\omega(r)$ is real giving
rise to complex eigenvalues of $h$. The matrix
elements of the Hamiltonian (\ref{hamil}) are evaluated in the quark
spinor basis proposed in ref.\cite{Ka84}. These spinors
are characterized by their grand spin eigenvalue $G$ and their
parity transformation properties. Since (\ref{hamil}) commutes with
the grand spin operator and is parity invariant as well, the Hamiltonian
(\ref{hamil}) is diagonalized for each grand spin and parity eigenvalue
separately. The momentum eigenvalues of the basis spinors are
discretized by putting the system into a finite spherical box of radius
$D$ and demanding the upper components to vanish at the boundary for
spinors with parity eigenvalue $(-1)^G$ and the lower components for
spinors with parity eigenvalue $(-1)^{G-1}$. We list all matrix elements
of operators appearing in (\ref{hamil}) as well as the explicit form of
the basis spinors in appendix A.  The boundary conditions of ref.\cite{Ka84}
have the advantage that there are no spurious contributions to the
equations of motion for the $\omega$ field stemming from the fact
that we only consider a finite basis in momentum space. The spurious
contributions to the profile function $G(r)$ of the $\rho$ field
appearing in the basis of ref.\cite{Ka84} are well under control and
may explicitly be eliminated. However, we will see later that there are
finite size effects for the $\omega$ meson profile originating from a
large but finite $D$.

For static configurations the eigenvalues of $\partial_\tau$,
$i\Omega_n=i(2n+1)\pi/T,$ with $(n=0,\pm1,$ $\pm2,..)$  may be separated
rendering the temporal part of the trace feasible\footnote{The
eigenfunctions of $\partial_\tau$ assume anti-periodic
boundary conditions in the Euclidean time interval $T$. The
$\Omega_n$ are the analogues of the Matsubara frequencies with $T$
figuring as inverse temperature.}.
{}~Thus the eigenvalues
$\lambda_{n,\nu}$ of the operator $\partial_\tau+h$ read:
\be
\lambda_{n,\nu}=-i\Omega_n+\epsilon_\nu=-i\Omega_n+
\epsilon_\nu^R+i\epsilon_\nu^I.
\label{lannu}
\ee
The fermion determinant is expressed in terms of the eigenvalues
$\lambda_{n,\nu}$:
\be
\A_R=\frac{1}{2}\sum_{\nu,n}{\rm log}(\lambda_{n,\nu}\lambda_{n,\nu}^*)
\qquad {\rm and}\qquad
\A_I=\frac{1}{2}\sum_{\nu,n}{\rm log}(\frac{\lambda_{n,\nu}}
{\lambda_{n,\nu}^*}).
\ee
Using (\ref{lannu}) the real part reads:
\be
\A_R&=&\frac{1}{2}\sum_{\nu,n}{\rm log}\big((\Omega_n-\epsilon_\nu^I)^2+
(\epsilon_\nu^R)^2\big) \nonumber \\
&&\qquad\to\ -\frac{1}{2}\sum_{\nu,n}\int_1^\infty
\frac{d\tau}{\tau} {\rm \exp}\big\{-\frac{\tau}{\Lambda^2}
\big((\Omega_n-\epsilon_\nu^I)^2+(\epsilon_\nu^R)^2\big)\big\}
\label{arprti}
\ee
according to the proper time regularization scheme (\ref{arreg}). For large
Euclidean time intervals ($T\rightarrow\infty$) the temporal part of
the trace may be performed
\be
\A_R=-\frac{T}{2}\sum_\nu\int_{-\infty}^{\infty}\frac{dz}{2\pi}
\int_1^\infty \frac{d\tau}{\tau}
{\rm \exp}\big\{-\frac{\tau}{\Lambda^2}\big(z^2
+(\epsilon_\nu^R)^2\big)\big\}
\label{arstatic}
\ee
where we have shifted the integration variable $z-\epsilon_\nu^I
\rightarrow z$. For $T\rightarrow\infty$ we may read off the
Dirac sea contribution to the real part of the energy functional
from $\A_R\rightarrow -TE_{\rm vac}^R$:
\be
E_{\rm vac}^R=\frac{N_C}{4\sqrt{\pi}}\sum_\nu |\epsilon_\nu^R|
\Gamma\big(-\frac{1}{2},(\epsilon_\nu^R/\Lambda)^2\big).
\label{ervac}
\ee
For the imaginary part we obtain
\be
\A_I=\frac{1}{2}\big(\sum_\nu\sum_{n=-\infty}^\infty
{\rm log}(\lambda_{\nu,n})-
\sum_\nu\sum_{n=-\infty}^\infty{\rm log}(\lambda_{\nu,n}^*)\big)=
\frac{1}{2}\sum_\nu\sum_{n=-\infty}^\infty{\rm log}
\frac{i\Omega_n-\epsilon_\nu}{i\Omega_n-\epsilon_\nu^*}
\ee
where we have reversed the sign in the first sum over the integer
variable $n$. Next we express $\A_I$ in terms of a parameter integral:
\be
\A_I=\frac{1}{2}\sum_\nu\sum_{n=-\infty}^\infty
\int_{-1}^1 d\lambda \frac{-i\epsilon_\nu^I}
{i\Omega_n-\epsilon_\nu^R-i\lambda\epsilon_\nu^I}.
\ee
In analogy to (\ref{ervac}) we may carry out the temporal trace in the
limit $T\rightarrow\infty$:
\be
\A_I=\frac{-i}{2}\sum_\nu \int_{-1}^1 d\lambda\, T
\int_{-\infty}^{\infty}\frac{dz}{2\pi}\frac{\epsilon_\nu^I}
{\big[i(z-\lambda\epsilon_\nu^I)-\epsilon_\nu^R\big]} .
\ee
Shifting the integration variable $z-\lambda\epsilon_\nu^I
\rightarrow z$ the integral over $\lambda$ may be done
\be
\A_I=\frac{-i}{2}\sum_\nu \epsilon_\nu^I
\int_{-\infty}^{\infty}\frac{dz}{2\pi}
\frac{-2\epsilon_\nu^R}{z^2+(\epsilon_\nu^R)^2}.
\label{3.14}
\ee
$\A_I$ is regularized in proper time by expressing the integrand as
a parameter integral:
\be
\frac{-1}{z^2+(\epsilon_\nu^R)^2}\to\int_{1/\Lambda^2}^\infty d\tau
{\rm exp}\big\{-\tau(z^2+(\epsilon_\nu^R)^2\big\}
\ee
which obviously is finite for $\Lambda\rightarrow\infty$. Continuing
the evaluation of $\A_I$ in analogy to eqs. (\ref{arprti}-\ref{ervac})
we find for the contribution of the Dirac sea to the imaginary part of the
Euclidean energy
$E_{\rm vac}^I$
\be
E_{\rm vac}^I=\frac{-N_C}{2}\sum_\nu \epsilon_\nu^I {\rm sign}
(\epsilon_\nu^R)\cases{1,&$\A_I\quad {\rm not}\ {\rm
regularized}$\cr
{\cal N}_\nu, &$\A_I\quad {\rm regularized}$\cr}
\label{eivac}
\ee
where
\be
{\cal N}_\nu = \frac{1}{\sqrt\pi}\Gamma\big(\frac{1}{2},
(\epsilon_\nu^R/\Lambda)^2\big)
\label{vacoccno}
\ee
are the vacuum ``occupation numbers" in the proper time regularization
scheme.  The upper case, of course, corresponds to the limit
$\Lambda\rightarrow\infty$. Obviously only the real part of the
one-particle energy eigenvalue is relevant for the regularization
of $\A_I$. Eqn. (\ref{eivac}) reveals that we have succeeded in finding a
regularization scheme for $\A_I$ that only involves quantities which
are strictly positive definite. This is not evident from
the definition of $\A_I$ (\ref{arai}).

For soliton configurations with vanishing $\omega$ (\ie
$\epsilon_\nu=\epsilon_\nu^R$) there is no contribution from the
imaginary part and eqn. (\ref{ervac}) is the expression
for the energy of the Dirac sea.

The total energy functional contains besides $E_{\rm vac}^R$ and
$E_{\rm vac}^I$ also the valence quark energy
\be
E_{\rm val}^R=N_C\sum_\nu\eta_\nu |\epsilon_\nu^R|\qquad
E_{\rm val}^I=N_C\sum_\nu\eta_\nu {\rm sign}(\epsilon_\nu^R)
\epsilon_\nu^I
\label{eeval}
\ee
with $\eta_\mu=0,1$ being the occupation numbers of the valence quark
and
anti-quark states.
Furthermore the meson energy is obtained by substituting the
{\it ans\"atze} (\ref{ansatz1}-\ref{ansatz3}) into (\ref{action}):
\be
E_{\rm m}&=&4\pi \int dr r^2\Bigl( m_\pi^2f_\pi^2(1- \cos\Theta(r))
\nonumber \\*
&& \qquad
+\big(\frac{m_\rho}{g_V}\big)^2\bigl(G^2(r)+\frac{1}{2}F^2(r)+F(r)H(r)+
\frac{3}{2}H^2(r)-2\omega^2(r)\bigr)\Bigr)
\label{emstatic}
\ee
Note that we are working in the isospin limit which implies
$m_\omega = m_\rho$.  Continuing back to Minkowski space we find for
the total energy functional:
\be
E[\Theta,\omega , G,F,H]
=E_{\rm val}^R+E_{\rm val}^I+E_{\rm vac}^R+E_{\rm vac}^I+E_m.
\label{efunct}
\ee

The equations of motion for the meson profiles are obtained by
extremizing the static Minkowski energy(\ref{efunct}). In a generic
way we may write:
\be
0=\frac{\delta E}{\delta\phi}&=&\frac{\delta E_m}{\delta\phi}
+\sum_{\kappa=R,I}
\sum_\mu \frac{\partial(E_{\rm val}^R+E_{\rm val}^I+E_{\rm vac}^R+
E_{\rm vac}^I)}{\partial \epsilon^\kappa_\mu}
\frac{\delta\epsilon^\kappa_\mu}{\delta\phi}
\label{eqm}
\ee
wherein $\phi$ denotes any of the meson profiles $\Theta,G,\omega,F$ or
$H$. Since $h$ is not Hermitean (in Euclidean space) we have to
distinguish between left and right eigenvectors of $h$. The
corresponding eigenvalue equations read:
\be
h|\Psi_\nu\rangle = \epsilon_\nu |\Psi_\nu\rangle\quad
\langle\tilde{\Psi}_\nu|h=\epsilon_\nu\langle\tilde{\Psi}_\nu|
\qquad \ie h^\dagger |\tilde{\Psi}_\nu\rangle = \epsilon_\nu^*
|\tilde{\Psi}_\nu\rangle.
\label{eigen}
\ee
The normalization condition is $\langle\tilde{\Psi}_\mu|\Psi_\nu\rangle =
\delta_{\mu\nu}$.  In order to evaluate the derivatives
${\delta\epsilon^\kappa_\mu}/{\delta\phi}$ it
is helpful to decompose the Hamiltonian operator
(\ref{hamil}) into Hermitean and anti-Hermitean parts
\be
h=h_\Theta+i\omega
\ee
where $h_\Theta$ includes all Hermitean terms of the Euclidean
Dirac Hamiltonian (\ref{hamil}). Obviously both, $h_\Theta$ and $\omega$,
are Hermitean implying $|\tilde{\Psi}_\nu\rangle=
|\Psi_\nu^*\rangle$. We may therefore extract the real and imaginary
parts of the one particle energy eigenvalue
\be
\epsilon_\nu^R&=&\frac{1}{2}\big(\langle\Psi_\nu^*|h|
\Psi_\nu\rangle+\langle\Psi_\nu|h^\dagger|\Psi_\nu^*\rangle\big)
\nonumber \\
&=&\langle\Psi_\nu^R|h_\Theta|\Psi_\nu^R\rangle
-\langle\Psi_\nu^I|h_\Theta|\Psi_\nu^I\rangle
-\langle\Psi_\nu^I|\omega|\Psi_\nu^R\rangle
-\langle\Psi_\nu^R|\omega|\Psi_\nu^I\rangle ,
\nonumber \\
\epsilon_\nu^I&=&\frac{1}{2}\big(\langle\Psi_\nu^*|h|
\Psi_\nu\rangle-\langle\Psi_\nu|h^\dagger|\Psi_\nu^*\rangle\big)
\nonumber \\
&=&\langle\Psi_\nu^R|\omega|\Psi_\nu^R\rangle
-\langle\Psi_\nu^I|\omega|\Psi_\nu^I\rangle
+\langle\Psi_\nu^I|h_\Theta|\Psi_\nu^R\rangle
+\langle\Psi_\nu^R|h_\Theta|\Psi_\nu^I\rangle
\label{erei}
\ee
where we employed the decomposition $|\Psi_\nu\rangle=|\Psi_\nu^R\rangle
+i|\Psi_\nu^I\rangle$. Note also that $\langle\Psi_\nu^*|=
\langle\Psi_\nu^R|+i\langle\Psi_\nu^I|$. We are now equipped with
expressions for the real and imaginary parts of the energy
eigenvalues $\epsilon_\mu$ which are suitable to evaluate the
derivatives with respect to the meson fields. For example we
have:
\be
\frac{\delta\epsilon^I_\nu}{\delta\omega(r)}=r^2
\int \frac{d\Omega}{4\pi} \left( \langle{\bf r}|\Psi_\nu^R\rangle
\langle\Psi_\nu^R|{\bf r}\rangle
-\langle{\bf r}|\Psi_\nu^I\rangle
\langle\Psi_\nu^I|{\bf r}\rangle\right).
\label{eqmom}
\ee
The expressions for the functional dependence of the energy eigenvalues
may now be substituted into the equations of motion (\ref{eqm}). The
individual equations for $\Theta, \omega, G, F$ and $H$ are displayed
in appendix B where we also discuss the spurious contributions to
the equation of motion for $G$.

At this point it is indispensable to explain the differences
between our approach and the treatment in ref. \cite{Doe93}.
We would like to stress that in order to regularize the
fermion determinant a continuation to Euclidean space has to be
performed. In Euclidean space the regularized fermion determinant is
well defined in terms of the complex energy eigenvalues
$\epsilon_\mu$ of the Euclidean Dirac Hamiltonian(\ref{hamil}).
This final expression for the fermion determinant may then be
continued back to Minkowski space yielding the energy
functional(\ref{efunct}). Due to the non-linear structure
of the regulator functions in (\ref{ervac}) and (\ref{eivac}) the
operations ``regularization" and ``continuation" do
${\underline {\rm not}}$ commute. However, this is just the
procedure used by the authors of ref. \cite{Doe93}: The
energy eigenvalues had firstly been continued to Minkowski
space and then the energy functional was regularized.
In appendix C we present a detailed comparison of the two different
treatments with the help of a toy model.

\vskip1cm
\stepcounter{chapter}
\leftline{\large \bf 4. Numerical Results}

\medskip

In this section we present the numerical results characterizing
the soliton solution. First we will explore the soliton
containing all meson fields as discussed in the preceding
sections. These results will then be compared to cases
where some of the vector meson fields are switched off
in order to examine the effects of various vector meson
fields on the soliton.

The meson profiles of the soliton solution are determined by
iteration. {\it I.e.}\ we start off with test profiles
for $\Theta,\omega, G, F$ and $H$ to evaluate the energy
eigenvalues $\epsilon_\mu$ of the static Hamiltonian (\ref{euham}).
Subsequently the meson profiles undergo modifications
according to the equations of motion (\ref{eqm}). In turn
these modified profiles serve as input for the static
Hamiltonian. This process is repeated until a convergent
solution is obtained, \ie a self-consistent meson field
configuration is constructed.

We have found stable self-consistent solutions for constituent
quark masses in the range 300MeV$\le M\le$400MeV\footnote{We do not
exclude the existence of stable solutions for an even larger range
of $M$ .}. In all these calculations the parameters have been kept
at their physical values (cf. section 2). The existence of these
stable solutions represents the main difference as compared to the
results of ref.\cite{Doe93}. Those authors only find solutions for
$m_\omega\ge870$MeV.

In table (4.1) we display the energy $E$ of the self-consistent
soliton solution. This table also contains the various
contributors to $E$ as they appear in eqn.(\ref{efunct})
for several values of the constituent mass $M$.
Furthermore we compare to the results obtained in the
chiral limit $(m_\pi=0)$.
\begin{table}
\tcaption{The soliton energy $E$ as well as its Dirac sea and mesonic
contributions $E_{\rm vac}$ and $E_{\rm m}$ for different values of the
constituent quark mass $M$. For the pion mass $m_\pi$ the physical
value (135MeV) and the chiral limit ($m_\pi=0$) are considered.
Also shown is the energy of the 'dived' level ($\epsilon_{\rm val}$).}
\newline
\centerline{\tenrm\smalllineskip
\begin{tabular}{||l|c|c|c|c|c|c||}
\hline
$M$ (MeV)    ~~~~~~~~~~   & \multicolumn{2}{c|} {300} &
 \multicolumn{2}{c|} {350}   & \multicolumn{2}{c||} {400} \\
\hline
$m_\pi$ (MeV)     &    0 &  135 &    0 &  135 &    0 & 135 \\
\hline
$E$ (MeV)         & 1117 & 1155 & 1024 & 1061 &  952 & 980 \\
\hline
$E^R_{\rm vac}$ (MeV) &  690 &  690 &  569 &  572 & 504 & 503 \\
\hline
$E^I_{\rm vac}$ (MeV) &   36 &   36 &   31 &   30 &   24 &  24 \\
\hline
$E_{\rm m}$ (MeV)   &  391 &  429 &  424 &  459 &  423 & 453 \\
\hline
$\epsilon^R_{\rm val}/M$     &-0.11 &-0.11 &-0.44 &-0.42 &-0.60 &-0.61\\
\hline
$\epsilon^I_{\rm val}/M$     & 0.10 & 0.10 & 0.09 & 0.09 & 0.07 & 0.08\\
\hline
\end{tabular}}
\end{table}

\begin{table}
\tcaption{Same as table (4.1) for $m_\pi=0$, however the imaginary part is
not regularized.}
\newline
\centerline{\tenrm\smalllineskip
\begin{tabular}{||l|c|c|c||}
\hline
$M$ (MeV)    ~~~~~~~~~~   & 300 & 350 & 400 \\
\hline
$E$ (MeV)         & 1227 & 1210 &   1175 \\
\hline
$E^R_{\rm vac}$ (MeV) &  669 &  563 &  460  \\
\hline
$E^I_{\rm vac}$ (MeV) &   55 &  257 &  298  \\
\hline
$E_{\rm m}$ (MeV)   &  335 &  389 &  417  \\
\hline
$\epsilon^R_{\rm val}/M$     & 0.02 &-0.28 &-0.47 \\
\hline
$\epsilon^I_{\rm val}/M$     & 0.13 & 0.16 & 0.16  \\
\hline
\end{tabular}}
\end{table}

The most striking result observed from table (4.1) is the fact
that the real part of the energy eigenvalue associated with
the valence quark state is negative! {\it I.e.}\ the valence quark
has joined the Dirac sea and thus the baryon number is completely
carried by the polarization of the Dirac sea. Thus the
soliton of the NJL model supports the picture of baryons as
topological solitons of meson fields. As a reminder we would like
to mention that for $\epsilon^R_{\rm val}<0$ the valence quarks'
contribution to the energy is already contained in $E^R_{\rm vac}$
and $E^I_{\rm vac}$. Thus it must not explicitly be added in
(\ref{eeval}). We also observe from table (4.1) that if the imaginary
part of the fermion determinant is regularized it only contributes a
minor part to the total energy. However, we see from table (4.2)
that this is no longer the case when the regularization of the
imaginary part is abandoned. In that case the total energy $E$
is almost independent of the constituent quark mass $M$ while
$E$ decreases by about $20\%$ in case the imaginary part is
regularized when $M$ is changed from 300MeV to 400MeV. For a
non-regularized imaginary part the valence quarks appear to be
slightly above the Dirac sea for $M$=300MeV. For all other cases the
valence quarks' energy remains negative, although less strongly bound
in the Dirac sea than for a regularized imaginary part where the
$\omega$ profile is less pronounced. Table (4.1) also reveals that
the inclusion of a finite pion mass does not alter the results for
the energy drastically. The total energy is increased by about
$40$MeV due to the additional term in $E_{\rm m}$ (\ref{emstatic}).

Next we wish to investigate the role of the different meson fields
for the NJL soliton\footnote{We do, however, not consider a system
containing the $\rho$ and $\omega$ mesons besides the pseudoscalar
fields.}. In order to do so we compare in table (4.3)
the results obtained for the soliton energy in cases with different
(axial-)vector mesons incorporated. Obviously, the inclusion of the
$\omega$ field always increases the soliton energy while the
$\rho$ and $a_1$ lower the energy. Though the latter result is
anticipated the former is somewhat surprising since the meson
profiles are obtained by extremizing the total energy. However,
this increase is understood by taking into account that the
$\omega$ field is proportional to the baryon number density which
in turn is constrained by unit baryon number. We also observe from
table (4.3) that the inclusion of any of the (axial-) vector
mesons lowers the energy eigenvalue of the valence quark. The
$a_1$ obviously effects the valence quark to join the Dirac
sea. This import result was already obtained previously\cite{Al92}
and the main conclusion to be drawn out of our present
calculation is the fact that this result is not spoiled by
the $\omega$ meson which was not included in ref.\cite{Al92};
on the contrary the $\omega$ meson gives an even stronger binding
of the valence quark. This may be understood by noting that
the $\omega$ is repulsive yielding a large spatial extension of
the soliton. This, in due, causes the valence quark energy to
drop.

\begin{table}
\tcaption{The soliton energy for various treatments of the NJL soliton. The
meson fields in the first line denote the allowed meson profiles. All numbers
are evaluated for a constituent quark mass $M=350$MeV and $m_\pi=0$. If
present, the imaginary part is regularized.}
\newline
\centerline{\tenrm\smalllineskip
\begin{tabular}{||l|c|c|c|c|c||}
\hline
 & $\pi$ & $\pi,\rho$ & $\pi,\omega$ & $\pi,\rho,a_1$ &  $\pi,\omega,\rho,a_1$
 \\ \hline
$E$ (MeV)         & 1214 &  957 & 1310 & 1010 & 1024 \\
\hline
$E^R_{\rm vac}$ (MeV) &  561 &  655 &  629 &  615 &  569 \\
\hline
$E^I_{\rm vac}$ (MeV) &    0 &    0 &  -56 &    0 &   31 \\
\hline
$E_{\rm m}$ (MeV)     &    0 &  155 &  -42 &  395 &  424 \\
\hline
$\epsilon^R_{\rm val}/M$     & 0.62 & 0.14 & 0.50 &-0.13 &-0.44\\
\hline
$\epsilon^I_{\rm val}/M$     &    0 &    0 & 0.24 &    0 & 0.09\\
\hline
\end{tabular}}
\end{table}

The repulsive character of the $\omega$ field may also be observed
directly from radial behavior of the chiral angle, $\Theta(r)$. In
fig. (4.1) we display $\Theta(r)$ for various treatments of the
NJL soliton. $\Theta(r)$ develops the largest tail in the case
when the $\omega$ meson field is the only one added to the chiral
field\cite{Al92a}. The axialvector field provides a significant
attraction resulting in a slope for the chiral field which is larger
than in the case when $\Theta(r)$ is the only field being present.
The inclusion of the $\omega$ meson on top of the isovector
mesons $\rho$ and $a_1$ alters the chiral angle only slightly.
A similar behavior can be observed for the $\rho$ meson profile
$G(r)$ (cf. fig. (4.2)) as well as the axialvector meson profiles
$F(r)$ and $H(r)$ (cf. fig. (4.3)). On the other hand the inclusion
of the isovector mesons on top of the $\pi-\omega$ system
significantly reduces the strength of the $\omega$ meson profile
as may be seen in fig. (4.4).

\begin{figure}
\vskip0.5cm
\fcaption{The chiral angle $\Theta(r)$ as a function of the
radial distance $r$. The solid line corresponds to the case when all
vector mesons are included; the dashed line to the $\pi-\omega$
system; the short dashed to the $\pi-\rho$ system; the
long dashed to the $\pi-\rho-a_1$ system and the dashed dotted to
the case when the chiral angle is the only field present.}
\end{figure}

\begin{figure}
\vskip0.5cm
\fcaption{The vector meson profile $G(r)$ as a function of the radial
distance $r$. Solid line: all vector meson fields are present;
short dashed denotes the $\pi-\rho$ system and the
long dashed to the $\pi-\rho-a_1$ system.}
\end{figure}

\begin{figure}
\vskip0.5cm
\fcaption{The axialvector meson profile $F(r)$ and $H(r)$
as functions of the radial distance $r$. $H(r)$ is non-vanishing
at the origin. The case when all vector mesons are present is
denote by the solid and dashed lines. The $\pi-\rho-a_1$ system
is represented by the long dashed and dashed dotted lines.}
\end{figure}

\begin{figure}
\vskip0.5cm
\fcaption{The vector meson profile $\omega(r)$ as a function of the radial
distance $r$. The solid line denotes the case when all vector
mesons are present while the dashed line represents the
$\pi-\omega$ system.}
\end{figure}

Of course, in the present form the soliton carries neither good
spin nor isospin quantum numbers, hence, it does not describe
physical baryon states. Following the work of Adkins, Nappi and
Witten\cite{Ad83} the soliton is projected onto baryon quantum
numbers by a cranking procedure which introduces time dependent
collective coordinates for the zero modes $R(t)$. Explicitly one
imposes the ansatz:
\be
\xi(\mbox{\boldmath $x$},t)=R(t)\xi(\mbox{\boldmath $x$})
R^{\dag}(t) \qquad R(t)\in SU(2).
\label{collrot}
\ee
and similarly for the isovector fields $\rho$ and $a_1$. Although
this approach has been successfully applied to the SU(2) NJL model of
pseudoscalar fields only\cite{Re89a} and even been generalized to
SU(3)\cite{We92a} this is not the whole story in the presence of
vector mesons. The collective rotation $R(t)$ excites additional vector
meson components as \eg the space components of $\omega$ and the time
components of $\rho$ which are absent in the static case\cite{Me87}.
Since the investigation of these excitation is beyond the scope of this
paper we will not discuss the resulting baryon spectrum, especially the
nucleon-$\Delta$ mass splitting. Nevertheless there are a  few
baryon properties which dependent only on the static fields and
may thus be evaluated at the present stage; the most prominent of
which is the axial charge of the nucleon, $g_A$.

In the NJL model the current field identities hold\cite{Eb86}. Therefore
the axial current $J_{5\mu}^i$ is directly proportional to the axial field:
\be
J_{5\mu}^i=-\frac{1}{4g_2}A^i_\mu
\label{j5}
\ee
wherein the superscript denotes the isospin component. Noting that
$g_A$ is obtained as the matrix element of $J_{5\mu}^i$ at zero
momentum transfer we immediately obtain
\be
g_A=-\frac{2\pi}{g_2}\int dr r^2\left[H(r)+\frac{1}{3}F(r)\right]
\langle D_{33}\rangle_p.
\label{ga}
\ee
$\langle D_{33}\rangle_p=1/3$ refers to the matrix element of
$\tr (R\tau_3R^\dagger\tau_3)$ between proton states. It is important
to mention that eqn. (\ref{ga}) represents an exact result which
is not subject to renormalization due to $\pi-a_1$ mixing.
Making use of the equation of motion for the axialvector profiles
$H(r)$ and $F(r)$ (see appendix B) we may reexpress $g_A$ as a mode
sum over quark spinors:
\be
g_A=-\frac{N_C}{3}\sum_\mu\Big\{&
\big[\langle\psi^R_\mu|\sigma_3\tau_3|\psi^R_\mu\rangle
-\langle\psi^I_\mu|\sigma_3\tau_3|\psi^I_\mu\rangle
+\langle\psi^R_\mu|\sigma_3\tau_3|\psi^I_\mu\rangle&
+\langle\psi^I_\mu|\sigma_3\tau_3|\psi^R_\mu\rangle\big]\eta_\mu
\nonumber \\
&+\big[\langle\psi^R_\mu|\sigma_3\tau_3|\psi^R_\mu\rangle
-\langle\psi^I_\mu|\sigma_3\tau_3\psi^I_\mu\rangle\big]
f_R\big(\epsilon_\mu/\Lambda)&\nonumber \\
&+\big[\langle\psi^R_\mu|\sigma_3\tau_3|\psi^I_\mu\rangle
+\langle\psi^I_\mu|\sigma_3\tau_3|\psi^R_\mu\rangle\big]
f_I\big(\epsilon_\mu/\Lambda)&
\Big\}
\label{gasum}
\ee
The regulator functions $f_{R,I}$ are listed in appendix B (cf.
eqns. (\ref{B3},\ref{B4})). The mode sum (\ref{gasum}) has the
advantage that it may be employed in models without axialvector
mesons.

\begin{table}
\tcaption{The axial charge $g_A$ of the nucleon in the various
treatments of the soliton in the NJL model. The last line
corresponds to the case when the imaginary part is not
regularized.}
\newline
\centerline{\tenrm\smalllineskip
\begin{tabular}{||l|c|c|c||}
\hline
$M$ (MeV)    ~~~~~~~~~   & 300  &  350   &  400 \\
\hline
$\pi$                    & ---  & 0.78  & 0.73 \\
\hline
$\pi,\omega$             & ---  & 0.98  & 1.03 \\
\hline
$\pi,\rho,a_1$           & 0.31 & 0.27 & 0.13 \\
\hline
$\pi,\omega,\rho,a_1$    & 0.34 & 0.25 & 0.23  \\
\hline
$\pi,\omega,\rho,a_1$    & 0.54 & 0.39 & 0.28  \\
\hline
\end{tabular}}
\end{table}

Unfortunately our numerical results which are listed in table (4.4)
are somewhat discouraging since they are well below the
numerical value $g_A=1.25$. This is especially pronounced in
case the valence quark energy is negative.
In order to understand the origin of this shortcoming let us
consider the case when the isoscalar field $\omega$ is
absent. Then eqn. (\ref{gasum}) reduces to
\be
g_A=-\frac{N_C}{3}\eta_{\rm val}\langle{\rm val}|\sigma_3\tau_3|
{\rm val}\rangle+\frac{N_C}{6}\sum_\mu{\rm sign}(\epsilon_\mu)
{\rm erfc}(|\epsilon_\mu|/\Lambda)\langle\mu|\sigma_3\tau_3|\mu\rangle
\label{gapi}
\ee
which is identical to the expression derived in ref.\cite{Wak90}.
Eqn. (\ref{gapi}) reveals that once the valence quark energy
has become negative its contribution to $g_A$ is strongly suppressed
by the proper time regularization. This suggests that a regularization
prescription which does not affect low-lying states as strongly
as the proper time scheme would be highly desirable in order to
describe $g_A$ correctly. This consideration is supported by a
simple modification of eqn. (\ref{gapi}). In the sum over all
eigenstates we replace the complementary error function by a
sharp cut-off function. Then the contributions from the low-lying
states are not affected by the regularization procedure. Choosing, \eg
a constituent mass of 300MeV the prediction for $g_A$ increases
drastically to 1.04 in the $\pi-\rho-a_1$ system.  Of course, this
exploration does not represent a consistent calculation but merely
demonstrates that $g_A$ strongly depends on the regularization description.
If we consider the $\pi-\omega$ system the strong repulsion of the
$\omega$ field transfers to an increased prediction for $g_A$. This is
also obvious from fig.(4.1) since in the absense of axialvector fields
$g_A$ may also be obtained from the size of the ``pion tail"\cite{Ad83}.

We may also investigate the isoscalar radius of the nucleon
without explicitly performing the collective quantization.
Again due to the current field identity the isoscalar charge
density is proportional to the $\omega$ profile yielding the
isoscalar radius:
\be
\langle r_{I=0}^2\rangle = {\rm N}^{-1}\frac{4m_\rho^2}{N_C g_V^2}
\int d^3r~ r^2~ \omega(r).
\label{rsqured}
\ee
The normalization factor
\be
{\rm N}=\frac{4m_\rho^2}{N_C g_V^2}\int d^3r~ \omega(r)
\ee
is one if the imaginary part of the determinant remains unregularized.

As already indicated in the previous section our numerical results
contain some spurious finite size contributions. These finite
size effects are manifested in a contribution to the $\omega$ meson
profile which is proportional to $D^{-3}$. Especially we find
that our numerical solution for $\omega(r)$ accquires a finite
value at the edge of the box: $\omega(r=D)\sim  D^{-3}$. From
eqn. (\ref{rsqured}) it may be observed that then
$\langle r_{I=0}^2\rangle$ would diverge as $D\rightarrow\infty$.
We eliminate this spurious contribution by enforcing $\omega(r)$
to vanish at the boundary. This is accomplished by including a
Lagrange multiplier $\lambda$ in the energy functional
(\ref{efunct}):
\be
E[\Theta,\omega , G,F,H]\longrightarrow E[\Theta,\omega , G,F,H]
+\lambda \int d^3r f_\epsilon(r)\omega^2(r)
\label{const}
\ee
wherein $f_\epsilon(r)$ is a positive radial function which vanishes
everywhere except within a small vicinity $\epsilon$ of $r=D$.
By iteration of the modified equations of motions $\lambda$ is
adjusted such that the additional term vanishes. Of course, there
is some arbitrariness in choosing $f_\epsilon(r)$. We demand the
change of the total energy with $\lambda$ included to deviate from
the case $\lambda=0$ by less than $1$MeV. In the region of
physical interest ($r\le2$fm) $\omega(r)$ remains almost unaltered
and we obtain the desired effect that $\langle r_{I=0}^2\rangle$
stays finite as $D\rightarrow\infty$ and assumes a constant value.
In case the $\omega$ field is not present we use the corresponding
unregularized mode sum (\ref{B6}) as input in eqn. (\ref{rsqured}).
The resulting data are displayed in table (4.5). The repulsive effect
of the isoscalar vector field $\omega$ is obvious and is not
completely compensated by the attraction provided by the isovector
fields $\rho$ and $a_1$. {\it I.e.} the prediction still
overestimates the experimental value $\langle r_{I=0}^2\rangle^{1/2}
\approx 0.8$fm.

\begin{table}
\tcaption{The isoscalar radius $\langle r_{I=0}^2\rangle^{1/2}$ of the
nucleon in various treatments of the NJL model. The radii are
given in fm.}
\newline
\centerline{\tenrm\smalllineskip
\begin{tabular}{||l|c|c|c||}
\hline
$M$ (MeV)    ~~~~~~~~~   & 300  &  350  &  400 \\
\hline
$\pi$                    & ---  & 0.89  & 0.76 \\
\hline
$\pi,\rho,a_1$           & 0.55 & 0.55  & 0.56 \\
\hline
$\pi,\omega$             & ---  & 2.06  & 1.77 \\
\hline
$\pi,\omega,\rho,a_1$    & 1.39 & 1.40  & 1.29 \\
\hline
\end{tabular}}
\end{table}

\vskip1cm
\stepcounter{chapter}
\leftline{\large \bf 5. Conclusions}

We have found stable soliton solutions in the NJL model with all
low-lying vector and axialvector meson fields included. The isoscalar
vector meson $\omega$ is excited only if the imaginary part of the
Euclidean action is taken into account. It is essential to note
that the necessary regularization can only be performed in
Euclidean space. This implies that the energy functional can be
defined properly only if one  continues to Euclidean space, regularizes
and then continues back to Minkowski space. This is quite distinct
from the procedure of ref. \cite{Doe93} where the one-particle
energy eigenvalues had firstly been continued to Minkowski space
and the resulting energy functional was ``regularized". As a
matter of fact, their results are quite different from ours.
Once they incorporate the $\omega$ meson they do not even find
stable soliton solutions for the physical value of the $\omega$ mass.

For reasonable values of the constituent quark mass we find that the
energy eigenvalue corresponding to the valence quark state is negative.
Therefore the baryon number is carried by the asymmetry of the
vacuum. Thus ${\underline{\rm the\ NJL\ model\ supports\ Witten's\
conjecture,}}$ ${\underline{\rm i.e.\ the\ Skyrmion\ picture\ of\
the\ baryon}}$. In this respect we remark that the $\omega$ meson is
even decreasing the valence quark level despite its repulsive nature.
This can be understood be noting that the $\omega$ meson broadens the other
meson profile functions thereby increasing the binding energy for the
valence quark state. However, if the imaginary part is not regularized
this effect is inverted.

The calculation of the isoscalar mean square radius of the nucleon is
plagued by a numerical deficiency stemming form the finite spatial
extension where we iterate the equations of motion. We have introduced a
Lagrange multiplier to keep the result for the isoscalar radius finite. The
prediction for the isoscalar mean square radius of the nucleon overestimates
the experimental value by about 40-50\%. Nevertheless we have seen the
repulsion provided by the isoscalar vector field as well as the
attraction due to the isovector (axial-) vector fields.

The axial coupling $g_A$ comes out much too small. We have demonstrated
that this behavior is due to a deficiency of the proper time
regularization which attaches ``weight factors" smaller than unity
already to levels with quite a small energy. We therefore conclude
that a regularization which does not effect low energy levels is
highly desirable.

In order to calculate other observables one has to project
the soliton solution on good spin and flavor states as was done for
the NJL soliton with pseudoscalars only in refs.\cite{Re89a,Goe91,Wak91}.
However, including the vector and axialvector mesons this is a
quite involved task since components which vanish on the classical
level but get excited by the collective rotation (\ref{collrot}) have
to be taken into account. On the other hand, we have seen that
the NJL soliton is close to a Skyrmion. Thus it makes sense to use
the self-consistent meson profiles in an action obtained by the
gradient expansion of the fermion determinant. A first step towards
this direction is to calculate static properties in this
approximation and compare those with the exact results.

Finally we would like to remark that problems arising from the
regularization can be overcome only if one takes a
renormalizable quark model as starting point. {\it E.g.} one
way were to use a bilocal effective quark interaction leading to
bilocal meson fields\cite{Ch85}. Whether such models allow for
soliton solutions and whether these solutions support
Witten's conjecture is, of course, an interesting question which
deserves further studies.

\medskip

\appendix
\raggedbottom

\vskip1cm
\stepcounter{chapter}
\leftline{\large \bf Appendix A: Matrix elements of the static Hamiltonian}

\medskip

For our numerical calculations we have used as basis the orthonormal
eigenstates of the free Dirac Hamiltonian
\be
h_0 = {\mbox{\boldmath $\alpha $}}{\bf p} + \beta M .
\ee
Since the static Hamiltonian commutes with the grand spin operator
$\mbox{\boldmath $G$}=\mbox{\boldmath $l$}+\mbox{\boldmath $\sigma$}/2
+\mbox{\boldmath $\tau$}/2$, our basis spinors are characterized by the
grand spin quantum number $G$ in addition to the total angular momentum
$j=G\pm1/2$ and the orbital angular momentum $l=G,G\pm1$. The latter
also determines the parity eigenvalue of the spinor, $\pi=(-1)^l$.
The grand spin eigenstates:
\be
\vert GMjl\rangle=\sum_{m_j,m_l}C^{GM}_{jm_j,1/2M-m_j}
C^{jm_j}_{lm_l,1/2m_j-m_l}\vert lm_l\rangle
\vert \frac{1}{2}m_j-m_l\rangle_S
\vert \frac{1}{2}M-m_j\rangle_I
\label{Geigenstates}
\ee
are two component spinors in both spin($S$)- and isospin($I$)-space.
The momentum, and therefore the energy,
is discretized by putting the system in a spherical box of radius $D$
and requiring appropriate boundary conditions. For the calculations
reported in this paper we used the boundary conditions of ref.\
\cite{Ka84}\footnote{For a discussion of a different type of boundary
condition see ref. \cite{We92a}.}. The discrete momenta $q^G_n$ are then
given by requiring
\be
 j_G(q^G_nD)=0 .
\label{A3}
\ee
where $j_G$ are the spherical Bessel functions. The energy eigenvalues
corresponding to the momenta $q_n$ in (\ref{A3}) are
$E^G_n=\pm\sqrt{(q^G_n)^2+m^2}$. Now the Dirac spinors are easily
constructed:
\vspace{.5cm}
\noindent

\vspace{.5cm}
\noindent
(i) $j=G\!+\!{1 \over 2};\ l=G$:
\be
\label{Spinor1}
   \vert G + + \ n \rangle &=& {\cal N}^{G}_{G+1n} \left( \begin{array}{rl}
  i w_n^{G+} j_G(q^G_nr)    & \vert G \ M \ (G\!+\!{1 \over 2})\ G \rangle   \\
  \noalign{\smallskip }
 w_n^{G-} j_{G+1}(q^G_nr) & \vert G \ M \ (G\!+\!{1 \over 2})\ (G\!+\!1)
                                              \rangle
              \end{array} \right)
\ee

\vspace{.3cm}
\noindent
(ii) $j=G\!-\!{1 \over 2};\ l=G$:
\be
\label{Spinor2}
  \vert G - + \ n \rangle &=& {\cal N}^{G}_{G+1n} \left( \begin{array}{rl}
   i w_n^{G+} j_G(q^G_nr)  & \vert G \ M \ (G\!-\!{1 \over 2})\ G \rangle  \\
  \noalign{\smallskip }
 -w_n^{G-} j_{G-1}(q^G_nr) &\vert G \ M \ (G\!-\!{1 \over 2})\ (G\!-\!1)
                                                  \rangle
              \end{array} \right)
\ee

\vspace{.3cm}
\noindent
(iii) $j=G\!+\!{1 \over 2};\ l=G+1$:
\be
  \vert G + - \ n \rangle &=& {\cal N}^{G}_{G+1n} \left( \begin{array}{rl}
   i  w_n^{G+} j_{G+1}(q^G_nr) & \vert G \ M \ (G\!+\!{1 \over 2})\ (G\!+\!1)
                                              \rangle   \\
  \noalign{\smallskip }
 -w_n^{G-} j_G(q^G_nr)     & \vert G \ M \ (G\!+\!{1 \over 2})\ G \rangle
              \end{array} \right)
\ee

\vspace{.3cm}
\noindent
(iv) $j=G\!-\!{1 \over 2};\ l=G-1$:
\be
  \vert G - - \ n \rangle &=& {\cal N}^{G}_{G+1n} \left( \begin{array}{rl}
   i  w_n^{G+} j_{G-1}(q^G_nr) & \vert G \ M \ (G\!-\!{1 \over 2})\ (G\!-\!1)
                                              \rangle   \\
  \noalign{\smallskip }
  w_n^{G-} j_G(q^G_nr)   & \vert G \ M \ (G\!-\!{1 \over 2})\ G \rangle
              \end{array} \right)
\ee
\noindent
where we used the following abbreviations
\be
 \label{KURZ}
 {\cal N}^{L}_{Gn} & = &D^{-{3\over 2}} \vert j_{G}(q^L_nD) \vert^{-1}
    \nonumber \\
 w_n^{L+} & = &\sqrt{1+{m \over E^L_n}} \nonumber \\
 w_n^{L-} & = &{\rm sign}\ (E^L_n)\sqrt{1-{m \over E^L_n}} .
\ee
We start with listing a few helpful relations involving the
two-component grand spin eigenstates (\ref{Geigenstates}) only:
\be
 \label{tau}
 \langle G M j_1 l_1 \vert {\mbox{\boldmath $\tau$}}\cdot{\bf \hat r}
 \vert G M j_2 l_2 \rangle
  = \left\{ \begin{array}{cl}
  \GK{1} & \mbox{for}\ j_1=j_2=G+\frac{1}{2} \ , \ \vert l_2-l_1 \vert = 1 \\
  \noalign{\medskip }
  \GK{-1} & \mbox{for}\ j_1=j_2=G-\frac{1}{2}\ , \ \vert l_2-l_1 \vert = 1 \\
  \noalign{\medskip }
  - \GK{\GS} & \mbox{for}\ j_1=G\pm\frac{1}{2},\ j_2=G\mp\frac{1}{2},\
\vert l_2-l_1 \vert = 1 \\
  \noalign{\medskip }
  0 & \mbox{otherwise}
        \end{array} \right.
\ee
\be
  \langle G M j_1 l_1 \vert (\mbox {\boldmath $\sigma $} \times {\bf \hat r})
 \mbox {\boldmath $\tau $}  \vert G M j_2 l_2 \rangle &=& i\ [
j_1(j_1+1)-l_1(l_1+1)  \\
   && -j_2(j_2+1)+l_2(l_2+1)]
  \langle G M j_1 l_1 \vert {\mbox{\boldmath $\tau$}}
    \cdot{\bf \hat r} \vert G M j_2 l_2 \rangle
   \nonumber
\ee
\be
\label{sig_r_tau_r}
 \langle G M j_1 l_1 \vert (\mbox {\boldmath $\sigma \cdot $} {\bf \hat r} )
  ({\mbox{\boldmath $\tau$}}\cdot{\bf \hat r})
   \vert G M j_2 l_2 \rangle
  = \left\{ \begin{array}{cll}
  \GK{-1} & \mbox{for} & j_1=j_2=G+\frac{1}{2} \ , l_1=l_2 \\
  \noalign{\medskip }
  \GK{1}  & \mbox{for} & j_1=j_2=G-\frac{1}{2} \ , l_1=l_2 \\
  \noalign{\medskip }
  \GK{\GS} & \mbox{for} & j_1=G\pm\frac{1}{2},\ j_2=G\mp\frac{1}{2},\ \\
   \noalign{\smallskip }
           &      & l_1=l_2=G \\   \noalign{\medskip }
  \GK{\GS} & \mbox{for} & j_1=G\pm\frac{1}{2},\ j_2=G\mp\frac{1}{2},\ \\
   \noalign{\smallskip }
           &      & \vert l_2-l_1 \vert = 2 \\  \noalign{\medskip }
  0 & \mbox{otherwise} &
        \end{array} \right.
\ee
and
\be
\label{sig_tau}
 \langle G M j_1 l_1 \vert (\mbox {\boldmath $\sigma \cdot \tau $} )
  \vert G M j_2 l_2 \rangle  = \left\{ \begin{array}{cl}
 1 & \mbox{for}\ j_1=j_2=G\pm\frac{1}{2} \ , l_1=l_2=G\pm 1 \\
  \noalign{\medskip }
  -\frac{2G-1}{2G+1}  & \mbox{for}\ j_1=j_2=G-\frac{1}{2} \ , l_1=l_2=G \\
  \noalign{\medskip }
  -\frac{2G-3}{2G+1}  & \mbox{for}\ j_1=j_2=G+\frac{1}{2} \ , l_1=l_2=G \\
  \noalign{\medskip }
  \GK{4 \sqrt{G(G+1)}} & \mbox{for}\ j_1=G\pm\frac{1}{2},
  \ j_2=G\mp\frac{1}{2},\ l_1=l_2=G \\
  \noalign{\medskip }
  0 & \mbox{otherwise}
        \end{array} \right.
\ee
A few helpful identities to verify the above expressions are:
\be
 (\mbox {\boldmath $\sigma \cdot \tau $} +1)^2 &=& 4 \\
 \left[ (\mbox {\boldmath $\sigma \cdot $} {\bf \hat r} )
({\mbox{\boldmath $\tau$}}\cdot{\bf \hat r})\right]^2 &=& {\bf 1} \\
 \left[\mbox {\boldmath $\sigma \cdot \tau $} -  (\mbox {\boldmath $\sigma
\cdot $} {\bf \hat r} ) ({\mbox{\boldmath $\tau$}}\cdot{\bf \hat r})
 \right]^2 &=& 2 - 2 (\mbox {\boldmath $\sigma \cdot $} {\bf \hat r} )
({\mbox{\boldmath $\tau$}}\cdot{\bf \hat r}) .
\ee

In the following we shall display all matrix elements which are needed
for the static Hamiltonian.

We introduce some additional abbreviations:
\be
 \begin{array}{rcl}
 a_{nm}^{ll'}&=&w_{n}^{G +}w_{m}^{G +}j_{l}(q^G_nr)j_{l'}(q^G_mr) \\
  \noalign{\medskip }
 b_{nm}^{ll'}&=&w_{n}^{G-}w_{m}^{G-}
   j_{l}(q^G_nr)j_{l'}(q^G_mr) \\
  \noalign{\medskip }
 c_{nm}^{ll'}&=&w_{n}^{G-}w_{m}^{G+}j_{l}(q^G_nr)j_{l'}(q^G_mr) \\
  \noalign{\medskip }
 N_{nm}&=& \vert j_{G+1}(q^G_nD) \vert^{-1}
                       \vert j_{G+1}(q^G_mD) \vert^{-1}
 \end{array}
\ee

\leftline{1. $\hat O=\beta f(r)$\footnote{We employ the
standard Dirac representation for the $\gamma$ matrices.}}
\be
 \begin{array}{rcl}
 \langle G + + \ n \ \vert \ \hat O\ \vert G + + \ m \rangle
  & = & \INT{(a_{nm}^{GG}-b_{nm}^{{G+1}{G+1}})}\\
  \noalign{\medskip }
 \langle G - + \ n \ \vert \ \hat O\ \vert G - + \ m \rangle
  & = & \INT{(a_{nm}^{GG}-b_{nm}^{{G-1}{G-1}})} \\
  \noalign{\medskip }
 \langle G + - \ n \ \vert \ \hat O\ \vert G + - \ m \rangle
  & = & \INT (a_{nm}^{{G+1}{G+1}}-b_{nm}^{GG})\\
  \noalign{\medskip }
 \langle G - - \ n \ \vert \ \hat O\ \vert G - - \ m \rangle
  & = & \INT(a_{nm}^{{G-1}{G-1}}-b_{nm}^{GG})
 \end{array}
\nonumber
\ee

\noindent

\leftline{2. $\hat O= i{\mbox{\boldmath $\tau$}}
\cdot{\bf \hat r}\beta\gamma_5 f(r)$}
\be
 \begin{array}{rcl}
 \langle G + + \ n \ \vert \ \hat O\  \vert G + + \ m \rangle
  & = & \INT{ \GK{1}(c_{nm}^{{G+1}G}+c_{mn}^{{G+1}G})}\\
  \noalign{\medskip }
 \langle G - + \ n \ \vert \  \hat O\ \vert G - + \ m \rangle
  & = & \INT{ \GK{1}(c_{nm}^{{G-1}G} +c_{mn}^{{G-1}G})}\\
  \noalign{\medskip }
 \langle G + + \ n \ \vert \ \hat O\ \vert G - + \ m \rangle
  & = & \INT \GK{\GS}(c_{mn}^{{G-1}G}-c_{nm}^{{G+1}G}) \\
  \noalign{\medskip }
 \langle G - + \ n \ \vert \ \hat O\ \vert G + + \ m \rangle
  & = & \INT \GK{\GS}(c_{nm}^{{G-1}G}-c_{mn}^{{G+1}G}) \\
  \noalign{\medskip }
 \langle G + - \ n \ \vert \ \hat O\ \vert G + - \ m \rangle
  & = & \INT{ \GK{-1}(c_{nm}^{G{G+1}}+c_{mn}^{G{G+1}})}\\
  \noalign{\medskip }
\end{array}
\nonumber
\ee
\be
\begin{array}{rcl}
 \langle G - - \ n \ \vert \ \hat O\ \vert G - - \ m \rangle
  & = & \INT{ \GK{-1}(c_{nm}^{G{G-1}}+c_{mn}^{G{G-1}})}\\
  \noalign{\medskip }
 \langle G + - \ n \ \vert \ \hat O\ \vert G - - \ m \rangle
  & = & \INT \GK{\GS}(c_{nm}^{G{G-1}}-c_{mn}^{G{G+1}})\\
  \noalign{\medskip }
 \langle G - - \ n \ \vert \ \hat O\ \vert G + - \ m \rangle
  & = & \INT \GK{\GS}(c_{mn}^{G{G-1}}-c_{nm}^{G{G+1}})
 \end{array}
\nonumber
\ee

\noindent

\leftline{3. $\hat O=  (\mbox {\boldmath $\alpha $} \times {\bf \hat r} )
{\mbox{\boldmath $\cdot\tau$}}f(r)$}
\be
 \begin{array}{rcl}
 \langle G + + \ n \ \vert \ \hat O\ \vert G + + \ m \rangle
  & = & \INT \GK{2(G+1)} (c_{nm}^{{G+1}G}+c_{mn}^{{G+1}G})\\
  \noalign{\medskip }
 \langle G - + \ n \ \vert \ \hat O\ \vert G - + \ m \rangle
  & = & \INT \GK{-2G} (c_{nm}^{{G-1}G}+c_{mn}^{{G-1}G}) \\
  \noalign{\medskip }
 \langle G + + \ n \ \vert \ \hat O\ \vert G - + \ m \rangle
  & = & \INT \GK{\GS}(c_{mn}^{{G-1}G}-c_{nm}^{{G+1}G})\\
  \noalign{\medskip }
 \langle G - + \ n \ \vert \ \hat O\ \vert G + + \ m \rangle
  & = & \INT \GK{\GS} (c_{nm}^{{G-1}G}-c_{mn}^{{G+1}G}) \\
  \noalign{\medskip }
 \langle G + - \ n \ \vert \ \hat O\ \vert G + - \ m \rangle
  & = & \INT \GK{2(G+1)} (c_{nm}^{G{G+1}}+c_{mn}^{G{G+1}})\\
  \noalign{\medskip }
 \langle G - - \ n \ \vert \ \hat O\ \vert G - - \ m \rangle
  & = & \INT \GK{-2G} (c_{nm}^{G{G-1}}+c_{mn}^{G{G-1}})\\
  \noalign{\medskip }
 \langle G + - \ n \ \vert \ \hat O\ \vert G - - \ m \rangle
  & = & \INT \GK{\GS} (c_{mn}^{G{G+1}}-c_{nm}^{G{G-1}})\\
  \noalign{\medskip }
 \langle G - - \ n \ \vert \ \hat O\ \vert G + - \ m \rangle
  & = & \INT \GK{\GS} (c_{nm}^{G{G-1}}-c_{mn}^{G{G-1}})
 \end{array}
\nonumber
\ee

\noindent

\leftline{4. $ \hat O= (\mbox {\boldmath $\sigma \cdot $} {\bf \hat r} )
(\mbox {\boldmath $\tau  \cdot $}  {\bf \hat r})f(r)$}

\be
 \begin{array}{rcl}
 \langle G + + \ n \ \vert \ \hat O\ \vert G + + \ m \rangle
  & = & \INT \GK{-1} (a_{nm}^{GG}+b_{nm}^{{G+1}{G+1}})\\
  \noalign{\medskip }
 \langle G - + \ n \ \vert \ \hat O\ \vert G - + \ m \rangle
  & = & \INT \GK{1} (a_{nm}^{GG}+b_{nm}^{{G-1}{G-1}})\\
  \noalign{\medskip }
 \langle G + + \ n \ \vert \ \hat O\ \vert G - + \ m \rangle
  & = & \INT \GK{\GS} (a_{nm}^{GG}-b_{nm}^{{G+1}{G-1}})\\
  \noalign{\medskip }
 \langle G - + \ n \ \vert \ \hat O\ \vert G + + \ m \rangle
  & = & \INT \GK{\GS} (a_{nm}^{GG}-b_{nm}^{{G-1}{G+1}}) \\
  \noalign{\medskip }
 \langle G + - \ n \ \vert \ \hat O\ \vert G + - \ m \rangle
  & = & \INT \GK{-1} (a_{nm}^{{G+1}{G+1}}+b_{nm}^{GG})\\
  \noalign{\medskip }
 \langle G - - \ n \ \vert \ \hat O\ \vert G - - \ m \rangle
  & = & \INT \GK{1} (a_{nm}^{{G-1}{G-1}}+b_{nm}^{GG})\\
  \noalign{\medskip }
 \langle G + - \ n \ \vert \ \hat O\ \vert G - - \ m \rangle
  & = & \INT \GK{\GS} (a_{nm}^{{G+1}{G-1}}-b_{nm}^{GG})\\
  \noalign{\medskip }
 \langle G - - \ n \ \vert \ \hat O\ \vert G + - \ m \rangle
  & = & \INT \GK{\GS} (a_{nm}^{{G-1}{G+1}}-b_{nm}^{GG})
 \end{array}
\nonumber
\ee

\noindent

\leftline{5. $ \hat O=(\mbox {\boldmath $\sigma\cdot\tau$})f(r)$}
\be
 \begin{array}{rcl}
 \langle G + + \ n \ \vert \ \hat O\ \vert G + + \ m \rangle
  & = & \INT{(- \GK{2G+3}a_{nm}^{GG}+b_{nm}^{{G+1}{G+1}})}\\
  \noalign{\medskip }
 \langle G - + \ n \ \vert \ \hat O\ \vert G - + \ m \rangle
  & = & \INT{(- \GK{2G-1}a_{nm}^{GG}+b_{nm}^{{G-1}{G-1}})}\\
  \noalign{\medskip }
 \langle G + + \ n \ \vert \ \hat O\ \vert G - + \ m \rangle
  & = & \INT{( \GK{4 \sqrt{G(G+1)}}a_{nm}^{GG})}\\
  \noalign{\medskip }
 \langle G - + \ n \ \vert \ \hat O\ \vert G + + \ m \rangle
  & = & \INT{( \GK{4 \sqrt{G(G+1)}}a_{nm}^{GG})} \\
  \noalign{\medskip }
\end{array}
\nonumber
\ee
\be
\begin{array}{rcl}
 \langle G + - \ n \ \vert \ \hat O\ \vert G + - \ m \rangle
  & = & \INT{(a_{nm}^{{G+1}{G+1}}-\GK{2G+3}b_{nm}^{GG})}\\
  \noalign{\medskip }
 \langle G - - \ n \ \vert \ \hat O\ \vert G - - \ m \rangle
  & = & \INT{(a_{nm}^{{G-1}{G-1}}-\GK{2G-1}b_{nm}^{GG})}\\
  \noalign{\medskip }
 \langle G + - \ n \ \vert \ \hat O\ \vert G - - \ m \rangle
  & = & \INT{(- \GK{4 \sqrt{G(G+1)}}b_{nm}^{GG})}\\
  \noalign{\medskip }
 \langle G - - \ n \ \vert \ \hat O\ \vert G + - \ m \rangle
  & = & \INT{(- \GK{4 \sqrt{G(G+1)}}b_{nm}^{GG})}
 \end{array}
\nonumber
\ee

\vskip1cm
\stepcounter{chapter}
\leftline{\large \bf Appendix B: Equations of motion for the meson profiles}

\medskip

In this appendix the equations of motion (\ref{eqm}) are displayed in
detail. The functional derivative of the one particle energies with
respect to the fields yields expressions involving the eigenfunctions
(\ref{eigen}). On the other hand,
the derivative of the total energy with respect to the
one particle energies results in regularization functions for the vacuum
part and the occupation number for the valence part. As the
eigenfunctions (\ref{eigen}) occur only in certain combinations in the
equations of motion it is convenient to define quark density matrices.

The quark scalar density matrix $\rho(\mbox{\boldmath $x,y$)}$ may be
decomposed into contributions due to valence and sea quarks:
\be
\rho(\mbox{\boldmath $x,y$)} & = & \rho_R^{\rm val} + \rho_I^{\rm val} +
\rho_R^{\rm vac} + \rho_I^{\rm vac},\nonumber \\*
\rho_R^{\rm val}(\mbox{\boldmath $x,y$}) & = & \sum_\nu\Big\{
\psi^R_\nu(\mbox{\boldmath $x$}) \bar
\psi^R_\nu(\mbox{\boldmath $y$})
-\psi^I_\nu(\mbox{\boldmath $x$}) \bar
\psi^I_\nu(\mbox{\boldmath $y$})\Big\}\eta_\nu ,\nonumber \\*
\rho_I^{\rm val}(\mbox{\boldmath $x,y$}) & = &  \sum_\nu\Big\{
\psi^R_\nu(\mbox{\boldmath $x$}) \bar
\psi^I_\nu(\mbox{\boldmath $y$})
+\psi^I_\nu(\mbox{\boldmath $x$}) \bar
\psi^R_\nu(\mbox{\boldmath $y$})\Big\} \eta_\nu ,\nonumber \\*
\rho_R^{\rm vac}(\mbox{\boldmath $x,y$}) & = & \sum_\nu\Big\{
\psi^R_\nu(\mbox{\boldmath $x$})\bar
\psi^R_\nu(\mbox{\boldmath $y$})
-\psi^I_\nu(\mbox{\boldmath $x$})\bar
\psi^I_\nu(\mbox{\boldmath $y$})\Big\}f_R\big(\epsilon_\nu/\Lambda),
\nonumber \\*
\rho_I^{\rm vac}(\mbox{\boldmath $x,y$}) & = &  \sum_\nu\Big\{
\psi^R_\nu(\mbox{\boldmath $x$})\bar
\psi^I_\nu(\mbox{\boldmath $y$})
+\psi^I_\nu(\mbox{\boldmath $x$})\bar
\psi^R_\nu(\mbox{\boldmath $y$})\Big\}f_I\big(\epsilon_\nu/\Lambda) .
\ee
Similarly, the quark number (or baryon number) density matrix
$b(\mbox{\boldmath $x,y$})$ is written as
\be
b(\mbox{\boldmath $x,y$})& = & b_R^{\rm val} + b_I^{\rm val} +
b_R^{\rm vac} + b_I^{\rm vac},\nonumber \\*
b_R^{\rm val}(\mbox{\boldmath $x,y$}) & = & \sum_\nu\Big\{
\psi^R_\nu(\mbox{\boldmath $x$})
\psi^{R\dagger}_\nu(\mbox{\boldmath $y$})
-\psi^I_\nu(\mbox{\boldmath $x$})
\psi^{I\dagger}_\nu(\mbox{\boldmath $y$})\Big\}
\eta_\nu, \nonumber \\*
b_I^{\rm val}(\mbox{\boldmath $x,y$}) & = & - \sum_\nu\Big\{
\psi^I_\nu(\mbox{\boldmath $x$})
\psi^{R\dagger}_\nu(\mbox{\boldmath $y$})
+\psi^R_\nu(\mbox{\boldmath $x$})
\psi^{I\dagger}_\nu(\mbox{\boldmath $y$})\Big\}
\eta_\nu, \nonumber \\*
b_R^{\rm vac}(\mbox{\boldmath $x,y$}) & = & \sum_\nu\Big\{
\psi^R_\nu(\mbox{\boldmath $x$})
\psi^{R\dagger}_\nu(\mbox{\boldmath $y$})
-\psi^I_\nu(\mbox{\boldmath $x$})
\psi^{I\dagger}_\nu(\mbox{\boldmath $y$})\Big\}
f_I\big(\epsilon_\nu/\Lambda), \nonumber \\*
b_I^{\rm vac}(\mbox{\boldmath $x,y$}) & = & - \sum_\nu\Big\{
\psi^I_\nu(\mbox{\boldmath $x$})
\psi^{R\dagger}_\nu(\mbox{\boldmath $y$})
+\psi^R_\nu(\mbox{\boldmath $x$})
\psi^{I\dagger}_\nu(\mbox{\boldmath $y$})\Big\}
f_R\big(\epsilon_\nu/\Lambda).
\ee
Note that the baryon density matrix
$b(\mbox{\boldmath $x,y$})$ differs from the scalar density
matrix $\rho (\mbox{\boldmath $x,y$})$ not only by the
additional factor $\gamma_0$ but also by an exchange
of the regulator functions $f_R$ and $f_I$ which are given as
the derivatives of the energy functional with respect to the
energy eigenvalue $\epsilon_\nu$:
\be
f_R\big(\epsilon_\nu/\Lambda)&=&\cases{-\frac{1}{2}
{\rm sign}(\epsilon_\nu^R){\cal N}_\nu,
&$\A_I\quad {\rm not}\ {\rm regularized},$\cr
&\cr
-\frac{1}{2}{\rm sign}(\epsilon_\nu^R)
{\cal N}_\nu+\frac{1}{\sqrt\pi}(\epsilon_\nu^I/\Lambda){\rm exp}
\big(-(\epsilon_\nu^R/\Lambda)^2\big),&$\A_I\quad {\rm regularized}$\cr}
\label{B3}
\ee
\be
f_I\big(\epsilon_\nu/\Lambda)&=&-\frac{1}{2}{\rm sign}(\epsilon_\nu^R)
\cases{1,&$\A_I\quad {\rm not}\ {\rm regularized}$\cr
&\cr {\cal N}_\nu
&$\A_I\quad {\rm regularized}$\cr}
\label{B4}
\ee
with the vacuum occupation number ${\cal N}_\nu$ being
defined in (\ref{vacoccno}).
With these definitions the equations of motion (\ref{eqm}) for the meson
profiles $\Theta , \omega , G,F$ and $H$ read
\be
\sin\Theta(r) &=&
\frac M {m_\pi^2f_\pi^2} N_c \tr\int \frac {d\Omega}{4\pi}\,
\Bigl( \sin\Theta(r) - i\gamma_5 {\mbox{\boldmath $\tau$}}\cdot{\bf
\hat r} \cos\Theta(r) \Bigr) \rho(\mbox{\boldmath $x,x$}) ,
\label{B5}
\\*
\omega(r)&=&\frac{g_V^2}{4m_\rho ^2}
N_c \tr \int \frac {d\Omega}{4\pi}\, b(\mbox{\boldmath $x,x$}) ,
\label{B6}
\\*
G(r) &=& - \frac{g_V^2}{4m_\rho ^2}
N_c \tr \int \frac {d\Omega}{4\pi}\, \Bigl(
({\mbox{\boldmath $\gamma $}} \times {\bf \hat r}) \cdot
{\mbox{\boldmath $\tau$}} \Bigr) \rho(\mbox{\boldmath $x,x$}) ,
\label{B7}
\\*
F(r) &=& - \frac{g_V^2}{4m_\rho ^2}
N_c \tr \int \frac {d\Omega}{4\pi}\, \beta \Bigl( 3 ({\mbox{\boldmath
$\sigma $}}\cdot {\bf \hat r}) ( {\mbox{\boldmath $\tau$}} \cdot {\bf
\hat r}) - ({\mbox{\boldmath $\sigma $}}\cdot {\mbox{\boldmath $\tau$}}
)\Bigr) \rho(\mbox{\boldmath $x,x$}) ,
\\*
H(r) &=& \frac{g_V^2}{4m_\rho ^2}
N_c \tr \int \frac {d\Omega}{4\pi}\, \beta \Bigl( ({\mbox{\boldmath
$\sigma $}}\cdot {\bf \hat r}) ( {\mbox{\boldmath $\tau$}} \cdot {\bf
\hat r}) - ({\mbox{\boldmath $\sigma $}}\cdot {\mbox{\boldmath $\tau$}} )\Bigr)
\rho(\mbox{\boldmath $x,x$}) .
\ee
The traces are over Dirac and isospin indices only. Note that
only the $\omega $ meson profile is given by a trace over the baryon
number density. All other meson profiles have as ``source" the scalar
quark density. Without regularization the difference between both
densities would have only been a factor $\gamma_0=\beta $. However, the
regularization makes the relation between the two densities, and
therefore the relation between the $\omega $ and the other meson
profiles, highly non--linear.

For the basis of ref.\cite{Ka84} the $RHS$ of eqn. (\ref{B7})
does not vanish for the vacuum configuration yielding some
spurious contributions to $G(r)$. These may, however, be
eliminated explicitly by subtracting the corresponding
value of the $RHS$ of eqn. (\ref{B7}) at each step of
the iteration procedure. These spurious contributions are
absent when the boundary conditions proposed in ref. \cite{We92a}
are employed at the expense of a non-vanishing
$RHS$ of eqn. (\ref{B6}) for the vacuum configuration. This
then would give similar spurious contributions to $\omega(r)$. Thus
we conclude that for the problem at hand both sets of
boundary conditions are equally well suited\footnote{This
equality is not the case when \eg matrix elements of
SU(3) generators are considered\cite{We92a}.}.

\vskip1cm
\stepcounter{chapter}\stepcounter{footnote}
\leftline{\large \bf Appendix C: Continuation to Euclidean space in a
toy model}

\medskip

In this appendix we discuss the differences of our approach\cite{Al92a}
and the treatment in \cite{Doe93} in the continuation prescription
to Euclidean space with the help of a $2\times2$ dimensional
toy model\footnote{We thank the anonymous referee of our earlier
paper \cite{Al92a} for bringing this toy model to our attention.}.

The non-Hermitean structure of our Hamiltonian (\ref{hamil}) is
already represented in the simple model:
\be
h=p \pmatrix{0&-i\cr i&0\cr}+i\omega_4\pmatrix{1&1\cr1&1\cr}.
\label{C1}
\ee
Comparing the structure of $h$ with a Dirac Hamiltonian we would like
to identify the variable $p$ with the ``momentum". Note that $\omega_4$
is real. The eigenvalues are
\be
\epsilon_{1,2}=i\omega_4\pm\sqrt{p^2-\omega_4^2}.
\label{C2}
\ee
Then the energy functional is given by (ignoring the subtraction
of the vacuum part, $\omega_4=0$):
\be
E^R_{\rm vac}={{N_C}\over{2\sqrt\pi}}\sqrt{p^2-\omega_4^2}
\Gamma\big(-{{1}\over{2}},(p^2-\omega_4^2)/\Lambda^2)\big)  \qquad
E^I_{\rm vac}=0
\label{C3}
\ee
for the case that $|p|>|\omega_4|$. Otherwise we have
from eqns. (\ref{arstatic}) and (\ref{eivac})\footnote{Consistency of
eqns. (\ref{3.14}) and (\ref{eivac}) enforces to define
${\rm sign}(0)=0$.}:
\be
E^R_{\rm vac}={{N_C}\over{4\sqrt\pi}}\Lambda \qquad
{\rm and} \qquad E^I_{\rm vac}=0
\label{C4}
\ee
\ie a constant energy functional and thus no contribution to
equation of motion.  Contrary to this, using the prescription
of ref.\cite{Doe93} yields a dynamical contribution of this mode:
\be
E^R_{\rm vac}={{N_C}\over{2\sqrt\pi}}\sqrt{p^2+\omega_0^2}
\Gamma\big(-{{1}\over{2}},(p^2+\omega_0^2)/\Lambda^2)\big)  \qquad
E^I_{\rm vac}=0
\label{C5}
\ee
For large ``momenta" $p$ eqn. (\ref{C5}) obviously is the continuation
of eqn (\ref{C3})  and
in both prescriptions the contribution of states with large $p$ is
suppressed via the regularization.  However, we see that for
$|p|<|\omega_4|$ the two approaches are no longer identical.
Especially, we see that in our prescription (\ref{C4}) the states with
small $p$ are not effected by the cut-off $\Lambda$ in contrast to
ref.\cite{Doe93} . This, of course, provides a physical motivation to
use our approach. It is also obvious from (\ref{C5}) that for
$|p|<|\omega_4|$ there is still a contribution to the equation of
motion in the formulation of ref.\cite{Doe93} in contradiction to
the formulation of the determinant in Euclidean space.

\end{document}